\newif{\ifjournal}
  \journaltrue

\ifjournal
  \documentclass[]{aa}
  \usepackage{graphicx,txfonts,amssymb,natbib}
  \renewcommand{\d}{\mathrm{d}}
  	
  \authorrunning{M. Maturi et al.}
  \titlerunning{An optimal filter for the detection of galaxy clusters
  through weak lensing}
\else
  \documentclass{paper}
\fi

\begin{document}

\title{An optimal filter for the detection of galaxy clusters through
  weak lensing}
\ifjournal
  \author{Matteo Maturi\inst{1,2}, Massimo Meneghetti\inst{1},
    Matthias Bartelmann\inst{1}, Klaus Dolag\inst{2,3}, Lauro
    Moscardini\inst{4}
    \institute{$^1$ ITA, Universit\"at Heidelberg,
    Albert-\"Uberle-Str.~2, 69120 Heidelberg, Germany\\
    $^2$ Dipartimento di Astronomia, Universit\`a di Padova, Vicolo
    dell'Osservatorio 2, 35120 Padova, Italy \\
    $^3$ Max-Planck-Institut f\"ur Astrophysik, P.O. Box 1523, 85740
    Garching, Germany\\
    $^4$ Dipartimento di Astronomia, Universit\`a di Bologna, Via
    Ranzani 1, 40127 Bologna, Italy}}
\else
  \author{Matteo Maturi$^{1,2}$, Massimo Meneghetti$^1$,
    Matthias Bartelmann$^1$, Klaus Dolag$^{2,3}$, Lauro
    Moscardini$^4$\\
    $^1$ ITA, Universit\"at Heidelberg, Albert-\"Uberle-Str.~2, 69121
    Heidelberg, Germany\\ 
    $^2$ Dipartimento di Astronomia, Universit\`a di Padova, Vicolo
    dell'Osservatorio 2, 35120 Padova, Italy\\
    $^3$ Max-Planck-Institut f\"ur Astrophysik, P.O. Box 1523, 85740
    Garching, Germany\\
    $^4$ Dipartimento di Astronomia, Universit\`a di Bologna, Via
    Ranzani 1, 40127 Bologna, Italy}
\fi
\date{\emph{Astronomy \& Astrophysics, submitted}}

\newcommand{\abstext}
  {We construct a linear filter optimised for detecting dark-matter
   halos in weak-lensing data. The filter assumes a mean radial
   profile of the halo shear pattern and modifies that shape by the
   noise power spectrum. Aiming at separating dark-matter halos from
   spurious peaks caused by large-scale structure lensing, we model
   the noise as being composed of weak lensing by large-scale
   structures and Poisson noise from random galaxy positions and
   intrinsic ellipticities. Optimal filtering against the noise
   requires the optimal filter scale to be smaller than typical halo
   sizes. Although a perfect separation of halos from spurious
   large-scale structure peaks is strictly impossible, we use
   numerical simulations to demonstrate that our filter produces
   substantially more sensitive, reliable and stable results than the
   conventionally used aperture-mass statistic.}

\ifjournal
  \abstract{\abstext}
\else
  \begin{abstract}\abstext\end{abstract}
\fi

\maketitle

\section{Introduction}

Are there dark halos in the universe, i.e.~dark-matter halos of
substantial mass which for some reason fail to produce light? This
question has recently attracted attention by the detection of
significant and apparently massive peaks in weak-lensing surveys which
are not seen to correspond to galaxy groups or clusters in the optical
or X-ray bands \citep[see e.g.][]{ER00.1,UM00.1,MI02.1,ER03.1}.

Dark-matter halos are embedded into large-scale structure. Measuring
the inhomogeneities of projected mass distributions, lensing
observations of halos are overlaid by the lensing signal of the
large-scale structure in front of and behind the halos. Being
approximately a Gaussian random field, lensing by large-scale
structure adds peaks and troughs to the signal which can be mistaken
for halos. It may thus be that part of the claimed dark-halo
detections are in fact peaks in the random weak gravitational lensing
signal of the large-scale structure.

Is there a way to separate the two types of lensing signal, and thus
to improve the reliability of weak-lensing halo detections? The
principal difficulty in doing so is the unsharp boundary between halos
and the large-scale structure. Nonetheless, we shall describe below
the construction of a linear matched filter which attempts to suppress
the weak-lensing signal from large-scale structures and yet leaves the
signal from sufficiently massive halos for detection.

Filtering in Fourier space is a flexible and powerful tool which has many
applications for galaxy cluster lensing, like the lensing on the Cosmic
Microwave Background radiation, where there is a strong superimposition of
different noise contributions (see e.g. \cite{SE00.1} and \cite{MAT04.1}),
and lensing on background sources. It could also be used for lensing on
background galaxies.

Our strategy is as follows. The large-scale structure can be
considered as composed of dark-matter halos of a broad and continuous
mass range. Yet, there is a characteristic length scale at each
cosmological epoch, the nonlinear scale, at which the variance of the
dark-matter density contrast becomes unity. This length scale
se
parates the small-scale regime where the dark-matter power spectrum
is dominated by the contributions of presumably virialised halos from
the large-scale regime on which the dark-matter density can be
considered as a linear superposition of linearly evolved perturbation
modes. In view of weak lensing, this suggests the operational
definition of large-scale structure lensing which is contributed by
the linearly evolved matter distribution beyond the non-linear scale,
and halo lensing, which is contributed by non-linear, gravitationally
bound, virialised structures.

We are aware that this operational definition may be questionable for
many applications of weak gravitational lensing. For instance, it is
well known that it is mainly the non-linear part of the dark-matter
power spectrum which has now been measured with spectacular success in
the weak-lensing surveys of many groups. Yet, we aim at a substantial
more modest goal here. Being confronted with peak detections in
weak-lensing data, we wish to know which of them can be attributed to
halos, and which of them are likely caused by the large-scale
structure.

In this attempt, we adopt the operational definition of the
distinction between large-scale structure and halo lensing, construct
a linear matched filter for halo lensing and model the lensing signal
by intervening large-scale structures as a noise component influencing
the shape of the matched filter. Matched filters for extracting
sources of interest embedded into substantial background noise were
previously proposed for many kind of observations, ranging from the
analysis of Cosmic Microwave Background maps
\citep{HN96.1,TE98.1,SA01.1,HE02.1,HE02.2,SA04.1} to weak lensing
tomography \citep{HE04.1}.

In Sect.~2, we briefly review those aspects of gravitational lensing
which are needed for the rest of the paper. We describe the filter
construction in Sect.~3 and contrast it with the more conventional
aperture mass in Sect.~4. Simulations described in Sect.~5 are used in
Sect.~6 to show that our suggested filter yields promising results. We
summarise and conclude in Sect.~7.

\section{Weak lensing quantities}
\label{sect:wlgeneral}

We briefly review in this section those gravitational-lensing concepts
which we shall use later. For a more detailed discussion, we refer the
reader to the review by \cite{BA01.1}. We start with an isolated lens
whose surface-mass density is $\Sigma(\vec\theta)$ at the angular
position $\vec\theta$ on the sky. Its lensing potential is
\begin{equation}
  \psi(\vec\theta)=
    \frac{4G}{c^2}\frac{D_{\rm l}D_{\rm s}}{D_{\rm ls}}
    \int\d^2\theta'\Sigma(\vec\theta')\,
    \ln\left|\vec\theta-\vec\theta'\right| \;,
\end{equation}
where $D_{\rm l,s,ls}$ are the usual angular-diameter distances
between the observer and the lens, the observer and the source, and
the lens and the source, respectively. The reduced deflection angle
experienced by a light ray crossing the lens plane at $\vec\theta$ is
the gradient of the potential,
\begin{equation}
  \vec\alpha(\vec\theta)=\vec\nabla\psi(\vec\theta) \;.
\label{eq:alpha}
\end{equation}
The image positions $\vec\theta$ for a source located at $\vec\beta$
are given by the lens equation
\begin{equation}
  \vec\theta=\vec\beta-\vec\alpha(\vec\theta) \;.
\end{equation}

For sources much smaller than typical scales on which the lens
properties vary, the lens mapping can be linearised. The deformation
of images with respect to the source is then given by the Jacobian
matrix
\begin{equation}
  A\equiv
  \frac{\partial\vec\beta}{\partial\vec\theta}=
  \left(\delta_{ij}-
  \frac{\partial^2\psi(\vec\theta)}{\partial\theta_i\partial\theta_j}
  \right)=\left(
  \begin{array}{cc}
    1-\kappa-\gamma_1 & -\gamma_2 \\
    -\gamma_2 & 1-\kappa+\gamma_1 \\
  \end{array} \right) \;.
\label{eq:jacobian}
\end{equation} 
Here, $\kappa$ is the convergence
\begin{equation}
  \kappa(\vec\theta)=\frac{\Sigma(\vec\theta)}{\Sigma_{\rm cr}}=
  \frac{1}{2}\left(\psi_{11}+\psi_{22}\right) \;,
\label{eq:k}
\end{equation}
i.e.~the surface-mass density scaled by its critical value
\begin{equation}
  \Sigma_{\rm cr}=
  \frac{c^2}{4\pi G}\frac{D_{\rm ls}}{D_{\rm l}D_{\rm s}} \;.
\end{equation}
The distortion is described by the two components of the shear, 
\begin{equation}
  \gamma_1=\frac{1}{2}\left(\psi_{11}-\psi_{22}\right)\;,\quad
  \gamma_2=\psi_{12} \;,
  \label{eq:g}
\end{equation}
which are combined into the complex shear $\gamma\equiv\gamma_1+{\rm
i}\gamma_2$. We further use the common abbreviation
\begin{equation}
  \frac{\partial^2 \psi(\vec{\theta})}{\partial \theta_i
    \partial \theta_j} \equiv \psi_{ij} \ .
\end{equation}

Outside critical curves, image ellipticities are determined by the
complex reduced shear
\begin{equation}
  g(\vec\theta)\equiv
  \frac{\gamma(\vec\theta)}{1-\kappa(\vec\theta)} \;.
\label{eq:rg}
\end{equation}
In the weak-lensing limit, $\kappa\ll1$, and the reduced shear
approximates the shear, $g\approx\gamma$, to first order.

Source and image shapes are quantified by the complex ellipticity
\begin{equation}
  \epsilon=\frac{a-b}{a+b}{\rm e}^{2{\rm i}\vartheta}\;,
\end{equation}
where $a$ and $b$ are the semi-major and semi-minor axes of an ellipse
fitting the object's surface-brightness distribution. The position
angle of the major axis of the ellipse is $\vartheta$. The expectation
value of the intrinsic source ellipticity $\epsilon_{\rm s}$ is
assumed to vanish.

A sufficiently small source with ellipticity $\epsilon_{\rm s}$ is
imaged to have an ellipticity
\begin{equation}
  \epsilon=\frac{\epsilon_{\rm s}+g}{1+g^*\epsilon_\mathrm{s}}\;,
\label{eq:ell}
\end{equation}
where the asterisk denotes complex conjugation. This equation
illustrates that the lensing distortion is determined by the reduced
shear, which is the only lensing quantity directly accessible through
measurements of galaxy ellipticities alone.

\section{Filter construction}

\subsection{General description}

We wish to construct a filter allowing the weak-lensing signal of
halos to be extracted from weak-lensing observations of wide
fields. We want the filter to suppress, as well as possible, the
weak-lensing signal of the large-scale structure which can otherwise
be mistaken as lensing by halos. As described in the introduction, we
are well aware that this separation cannot fully succeed because of
the unsharp boundary between lensing by large-scale structure and
dark-matter halos. According to the operational definition of
large-scale structure lensing given there, we shall model the weak
lensing by linearly evolved structures as a noise component to be
filtered out.

Consider the weak gravitational lensing signal of a dark-matter halo
with amplitude $A$ and angular shape $\tau(\vec\theta)$. The measured
signal $S(\vec{\theta})$ is contaminated by some noise
$N(\vec{\theta})$,
\begin{equation} \label{eqn:signal}
  S(\vec{\theta})=A\tau(\vec\theta)+N(\vec\theta)\;,
\end{equation}
whose components are to be specified below.

In our application, the observed signal is provided by the measured
ellipticities of the background galaxies, which are weakly distorted
by the foreground halo and the intervening large-scale structure. The
expectation value of the image ellipticities at a given position on
the sky is the reduced shear,
\begin{equation}
  \langle\epsilon\rangle=g\;.
\end{equation}

There are several noise contributions. First, the gravitational shear
is being measured at the random positions of the background
galaxies. This adds shot noise proportional to the number density of
the galaxies. Second, these galaxies are intrinsically elliptical,
thus the determination of a single galaxy ellipticity is a very noisy
measurement of the shear only. This adds a noise component
proportional to the variance of the intrinsic galaxy
ellipticity. Finally, there is the noise component which we are
primarily aiming at, which is caused by weak lensing of intervening
large-scale structures. This third source of noise will be modelled by
the linear dark-matter power spectrum, according to our operational
separation between halo and large-scale structure lensing.

These noise components are assumed to be random with zero mean and
isotropic such that their statistical properties are independent of
the position on the sky. These assumptions are well justified, since
in first approximation the background galaxies are randomly positioned
and oriented, and weak lensing by the large-scale structures is well
described by an isotropic Gaussian random field. The variances of the
noise components are conveniently described in the Fourier domain,
where their correlation functions are
\begin{equation}
  \langle\hat N(\vec k')^*\hat N(\vec k)\rangle=
  (2\pi)^2\,\delta(\vec k'-\vec k)P_N(k)\;;
\end{equation} 
the hats above symbols denote the Fourier transforms.

We wish to construct a linear filter $\Psi(\vec\theta)$ which, when
convolved with the signal, will yield an estimate $A_\mathrm{est}$ for
the amplitude $A$ at the position $\vec\theta$,
\begin{equation}
  A_{\rm est}(\vec\theta)=\int S(\vec\theta')
  \Psi(\vec\theta-\vec\theta')\,\d^2\theta'\;.
\label{eq:estimate}
\end{equation}  

We further want the filter to satisfy two constraints. First, we
require it to be unbiased, i.e.~the average error,
\begin{equation}
  b\equiv\langle A_{\rm est}- A\rangle=A\left[
    \int\Psi(\vec\theta)\tau(\vec\theta)\d^2\theta-1
  \right]
\end{equation}
has to vanish. Second, the measurement {\em noise} $\sigma$,
determined by the mean-squared deviation of the estimate from its true
value,
\begin{equation}
  \sigma^2\equiv\left\langle(A_{\rm est}-A)^2\right\rangle=
  b^2-\frac{1}{(2\pi)^2}\int\left|\hat\Psi(\vec{k})\right|^2\,
  P_N(k)\d^2 k\;,
\end{equation}
has to be minimal.

For finding a filter $\Psi$ satisfying these two conditions, we
combine them by means of a Lagrangian multiplier $\lambda$, carry out
the variation of $L=\sigma^2+\lambda b$ with respect to $\Psi$ and
thus find the function $\Psi$ minimising $L$. The solution of this
variational minimisation is given by
\begin{equation}
  \hat\Psi(\vec k)=\frac{1}{(2\pi)^2}\left[
    \int\frac{|\hat\tau(\vec k)|^2}{P_N(k)}\d^2k
  \right]^{-1}\,\frac{\hat\tau(\vec k)}{P_N(k)}\;.
\label{eq:optfilter}
\end{equation}
This is the usual result showing that the optimal linear filter shape
is determined by the shape $\tau$ of the signal, divided by the noise
power spectrum. The filter is thus made most sensitive for those
spatial frequencies where the signal $\hat\tau$ is large and the noise
$P_N(k)$ is small.

\subsection{Modelling the signal}

Before we can apply this formalism to the observed data, we need to
assume a model for the spatial distribution of the signal and a power
spectrum describing the noise properties of the data. Concerning the
signal, we assume that clusters are on average axially
symmetric. Thus, $\tau(\vec\theta)=\tau(|\vec\theta|)$. Specifically,
we use the density profile found in numerical simulations by
\cite{NA97.1} for modelling the average three-dimensional density
profile of dark-matter halos,
\begin{equation}
  \rho(r)=\frac{\rho_{\rm s}}{(r/r_{\rm s})(1+r/r_{\rm s})^2} \;,
\end{equation}
where $\rho_{\rm s}$ and $r_{\rm s}$ are characteristic density and
distance scales, respectively. These two parameters are not
independent, but stochastically related to a single parameter, which
can be chosen as the cluster mass.

The gravitational lensing properties of such a mass distribution have
been widely explored \citep[see e.g.][]{BA96.1,ME03.1}. Its lensing
potential is given by
\begin{equation}
  \psi(x)=4\kappa_{\rm s} h(x)\;,
\label{eq:nfwPsi}
\end{equation}
with
\begin{equation}
  h(x)=\frac{1}{2}\ln^2\frac{x}{2}+\left\{
    \begin{array}{r@{\quad\quad}l}
      2\,\mbox{arctan}^2\sqrt{\frac{x-1}{x+1}}  & (x>1) \\
     -2\,\mbox{arctanh}^2\sqrt{\frac{1-x}{1+x}} & (x<1) \\
      0 & (x=1)
    \end{array}\right.\;,
\end{equation}  
where $x$ is the projected distance from the lens centre in units of
the distance scale, $x=r/r_{\rm s}$. The reduced shear $g$ is obtained
from Eq.~(\ref{eq:nfwPsi}) using Eqs.~(\ref{eq:k}) and
(\ref{eq:g}). Its Fourier transform is defined as
\begin{equation}
  \hat\tau(\vec k)\equiv\hat{g}(\vec k)=\int\d^2x\,
  g(\vec x)\exp(\mathrm{i}\vec x\cdot\vec k)\;.
\end{equation}

\subsection{Noise power-spectrum}
\label{sect:noises}

As discussed earlier, we take three sources of noise into
account. These are the weak-lensing signal due to the large-scale
structure, the shot noise from the finite number of randomly placed
galaxies, and the noise due to the random intrinsic shape and
orientation of the sources.

The weak gravitational lensing effect of the large-scale structures
between the observer and the sources at a comoving distance $w$,
observed in the direction $\vec\theta$ on the sky, gives rise to an
effective convergence \citep[see e.g.][]{BA01.1}
\begin{eqnarray} 
  \kappa_{\rm eff}(\vec\theta,w)=
  \frac{3H_0^2\Omega_{\rm m}}{2c^2}\int_0^w\d w'
  \frac{f_K(w-w')f_K(w')}{f_K(w)}
  \frac{\delta[f_K(w')\vec\theta, w']}{a(w')}\;.
\label{eq:klss}
\end{eqnarray}
The factor out front contains the present-day matter-density parameter
$\Omega_\mathrm{m}$ and the Hubble constant $H_0$. The comoving
angular-diameter distance
\begin{equation}
  f_K(w)=\left\{
  \begin{array}{l@{\quad\quad}l}
    K^{-1/2}\sin(K^{1/2}w) & K>0 \\
    w & K=0 \\
    (-K)^{-1/2}\sinh(K^{1/2}w) & K<0
  \end{array}\right.
\end{equation}
is parametrised by the spatial curvature $K$. The density contrast
$\delta=(\rho-\bar\rho)/\bar\rho$ is taken at perpendicular distance
$f_K(w')\vec\theta$ and parallel distance $w'$, and $a(w')$ is the
scale factor of the universe at distance $w'$.

When the sources are distributed in comoving distance, $\kappa_{\rm
eff}(\vec\theta,w)$ needs to be averaged over the source-distance
distribution $G(w)$,
\begin{equation}
  \bar\kappa_{\rm eff}(\vec\theta)=\int_0^{w_{\rm H}}\d w G(w)
  \kappa_{\rm eff}(\vec\theta,w)\;,
\end{equation}
where the upper integration limit $w_{\rm H}$ is the comoving distance
to the horizon.

The power spectrum of the effective convergence $P_\kappa(k)$ is
related to the power spectrum of the three-dimensional density
fluctuations $P_\delta(k)$ by Limber's equation,
\begin{equation}
  P_\kappa(k)=\frac{9H_0^2\Omega_\mathrm{m}^2}{4 c^2}\,
  \int_0^{w_{\rm H}}\d w\,
  \frac{\bar{W}^2(w)}{a^2(w)}P_\delta\left(\frac{k}{f_K(w)},w\right)\;,
\label{eq:lss}
\end{equation}
in which the weight function $\bar W(w)$ is given by a line-of-sight
integral over distance ratios,
\begin{equation}
  \bar W(w)=\int_w^{w_{\rm H}}\d w'\,
  G(w')\,\frac{f_K(w'-w)}{f_K(w')}\;.
\end{equation}
The cosmic-shear and effective-convergence power spectra are identical,
\begin{equation}
  P_\gamma(k)=P_\kappa(k)\;.
\end{equation}

The noise contributions from random galaxy positions and ellipticities
are both modelled as Poisson noise, which has a flat power
spectrum. The shot-noise power spectrum from the galaxy positions is
given by the number density $n_\mathrm{g}$ of suitable galaxies,
\begin{equation}
  P_\mathrm{p}(k)\propto\frac{1}{n_\mathrm{g}}\;.
\end{equation}
The noise component caused by the random intrinsic shape and
orientation of the background galaxies has a power spectrum whose
amplitude depends on the variance $\sigma_{\epsilon_\mathrm{s}}^2$ of
the intrinsic ellipticity distribution,
\begin{equation}
  P_{\epsilon_{\rm s}}\propto\sigma_{\epsilon_{\rm s}}^2\;.
\end{equation}
The two noise components are combined into one power spectrum which
quantifies the noise of an ellipticity measurement,
\begin{equation}
  P_\epsilon(k)=\frac{\sigma_{\epsilon_\mathrm{s}}^2}{n_\mathrm{g}}\;.
\end{equation}
With the remaining noise components being independent, the total noise
power spectrum is the sum of the contributions,
\begin{equation}
  P_N(k)=P_\gamma(k)+P_\epsilon(k)\;.
\end{equation}

\begin{figure}[ht]
  \includegraphics[width=\hsize]{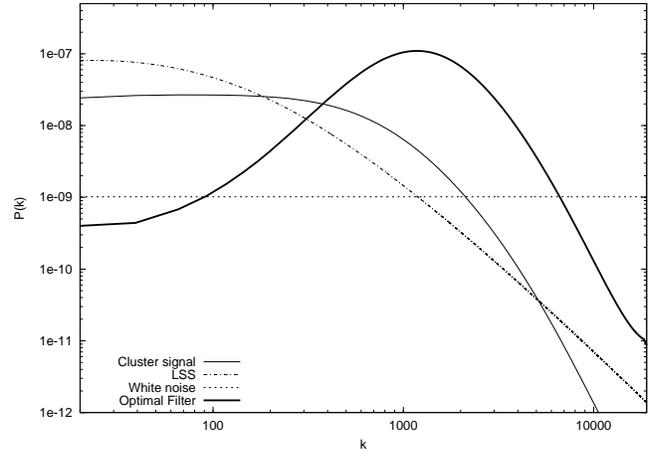}
\caption{Power spectra of the various components contributing to the
  definition of the optimal weak-lensing halo filter: effective
  convergence (dashed-dotted line), white noise composed of shot noise
  from the random galaxy positions and their intrinsic ellipticities
  (dotted line), and the model signal (solid line). The resulting
  filter is shown by the heavy solid line.}
\label{fig:1}
\end{figure} 

The contributions affecting the construction of the optimal filter are
shown in Fig.~\ref{fig:1}. The solid line shows the power spectra of
the signal, assuming an NFW lens model of $M=5\times
10^{14}\,h^{-1}\,M_\odot$ at redshift $z_{\rm l}=0.3$ and a source
redshift $z_{\rm s}=2$. The effective shear power spectrum and the
Poisson noise are given by the dash-dotted and the dotted lines,
respectively, where the latter contains the contributions from the
random galaxy positions and their random ellipticities. The bold solid
line shows the filter resulting from
Eq.~(\ref{eq:optfilter}). Unfortunately, but expectedly, the
large-scale structure power is significant on those spatial
frequencies where also most of the signal is located. This implies
that it will be impossible to perfectly separate the true signal from
the large-scale structure contaminations on the same scales. On the
other hand, the lensing noise contributed by the large-scale
structures on scales larger than the typical scale where the signal
power spectrum becomes negligible will be effectively filtered away.

\section{Aperture mass}
\label{sect:map}

For detecting halos through their weak-lensing signal, the aperture
mass statistic \citep[$M_{\rm ap}$,][]{SC96.2} is frequently used. Due
to its favourable statistical properties, it has turned into one of
the prime tools for analysing cosmic-shear statistics. Its application
to the detection of the weak-lensing patterns from dark-matter halos
has been shown by several authors \citep[see
e.g][]{ER00.1,SC04.2}.

The aperture mass is a weighted integral of the local surface mass
density,
\begin{equation}  
  M_{\rm ap}(\vec\theta)=\int\d^2\theta'\,\kappa(\vec\theta')
  U(\vec\theta-\vec\theta')\;,
  \label{eq:map}
\end{equation}
where $U(\vec\theta)$ is the weight function. If the aperture is
chosen to be circular, $U$ is axially symmetric,
$U(\vec\theta)=U(\theta)$. Since the convergence is not directly
measurable, Eq.~\ref{eq:map} is conveniently rewritten in terms of the
tangential shear $\gamma_\mathrm{t}$ as
\begin{equation}  
  M_{\rm ap}(\vec\theta)=\int\d^2\theta'\,\gamma_{\rm t}(\vec\theta')
  Q(\vec\theta-\vec\theta')\;,
  \label{eq:mapt}
\end{equation}
provided the filter $U$ is {\em compensated}, i.e.~satisfies
\begin{equation}
  \int_0^\theta\d\theta'\,\theta'\,U(\theta')=0\;.
\end{equation}
The function $Q(\theta)$ is related to $U(\theta)$ by
\begin{equation}
  Q(\theta)\equiv\frac{2}{\theta^2}\int_0^\theta\d\theta'\,
  \theta'\,U(\theta')-U(\theta)\;.
\end{equation}
The noise of the aperture mass is given by its variance
\begin{equation}
  \sigma^2_{M_{\rm ap}}=
  \frac{\pi\sigma_{\epsilon_\mathrm{s}}^2}{n_g}\,
  \int_0^\theta\d\theta'\,\theta'Q^2(\theta')\;.
\end{equation}
\citep{SC96.2}. 
 
The function $U$ can be chosen such to maximise the signal-to-noise
ratio $M_{\rm ap}/\sigma_{M_{\rm ap}}$. \cite{SC98.2} show that this
is the case if $Q$ mimics the tangential shear profile of the
lens. For example, \cite{SC04.2} use a function which approximates the
expected shear profile of a NFW halo.

\begin{figure}[ht]
  \includegraphics[width=\hsize]{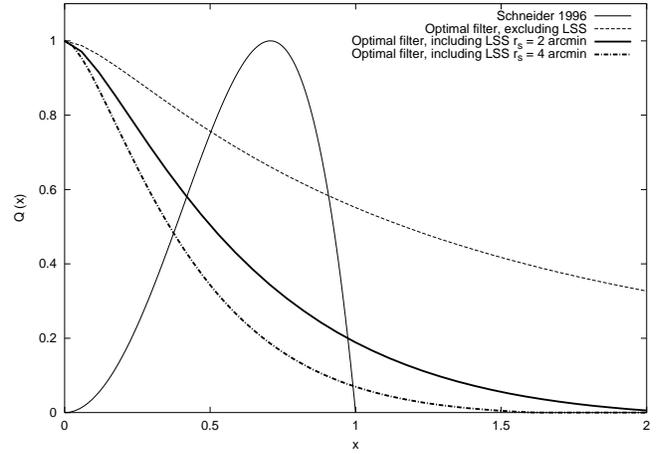}
\caption{Comparison between the filter profiles. Different lines refer to the
  conventional aperture mass (solid line), the optimal filter described in
  this paper but without the large-scale structure contribution to the noise
  power spectrum (dashed line), the full optimal filter for $r_s=2'$ and
  $r_s=4'$ (heavy solid heavy dot-dashed lines, respectively). The large-scale
  structure power spectrum modifies the filter shape according to the filter
  scale, by lowering its sensitivity on large scales where weak lensing by
  large-scale structure dominates.}
\label{fig:2}
\end{figure}

Thus, an assumption on the expected profile of the tangential shear
needs to be made in the construction of both the optimal filter and
the aperture mass. In this respect, the two functions $U$ and $\tau$
should be chosen to look very similar. Two differences are important,
however. First, the weight function $U$ used in the definition of the
aperture mass is compensated, while $\tau$ does not need to be. The
reason is that the aperture mass was specifically defined such as to
be proportional to the projected mass enclosed by the aperture. Our
goal here is more modest as we only ask for significant halo
detections, rather than for a determination of halo parameters apatr
from the overall shear amplitude. This allows us more freedom in the
choice of $\tau$ compared to $U$. Second, the shape of the optimal
filter differs from the profile function $\tau$ because it is modified
by the shape of the total noise power spectrum. A comparison between
the optimal filter and the function $Q(\theta)$ used by \cite{SC96.2}
is shown in Fig.~\ref{fig:2}. The solid and the dot-dashed lines refer
both to the optimal filter but for two different scale radii used for
modelling the signal. We will show in the following sections how the
two filters perform when applied to simulated data.
 
\section{Simulated source catalogues}

Numerical simulations are a unique tool for assessing the performance
of the optimal filter, individually and in comparison to the aperture
mass. In the following sections we describe the lens models and the
numerical methods used for producing simulated catalogues of weakly
distorted sources, subsequently analysed with our filter and the
aperture-mass statistic.

\subsection{Numerical models}

The numerical model used here is a simulation of a super-cluster like
region containing six halos with virial mass larger than $5.6 \times
10^{13}\,h^{-1}\,M_\odot$. They are embedded in an over-dense region
containing several less massive halos. The region is a section from an
existing dark-matter only large-scale cosmological simulation, later
re-simulated at higher resolution using the ``Zoomed Initial
Conditions'' (ZIC) technique \citep{TO97.2}. The parent simulation was
an $N$-body run of a box of $479\,h^{-1}\,\mathrm{Mpc}$ side length
containing $512^3$ particles of mass
$6.8\times10^{10}\,h^{-1}\,M_\odot$. The background cosmological model
is spatially flat with $\Omega_\mathrm{m}=0.3$ and
$\Omega_\Lambda=0.7$ at the final epoch, identified with redshift
zero \citep{YO01.1,JE01.1}.
 
From the output of this large cosmological simulation at redshift
zero, we selected the super-cluster region for the following
re-simulation. New initial conditions were constructed as
follows. First, the particles contained in the selected region were
traced back to their initial (Lagrangian) positions. These define a
region in Lagrangian space. Then, the initial density field in that
Lagrangian region was re-sampled by placing a higher number of
particles than were originally present, adding additional small-scale
power appropriately. Gas was introduced into the high-resolution
region by splitting each parent particle into a gas and a dark-matter
particle. The final mass resolution of these simulation was $m_{\rm
DM}=1.13\times 10^9\,h^{-1}M_\odot$ and $m_{\rm gas}=1.7\times
10^8\,h^{-1}M_\odot$ for dark matter and gas within the
high-resolution region, respectively. Thereby the simulation hosts a
usable region of $25\,h^{-1}$ Mpc.
   
Finally, the region was evolved down to redshift $z=0$ using the code
{\small GADGET-2}, a new version of the parallel tree-SPH simulation
code {\small GADGET} \citep{SP01.1}. The simulation was carried out
including only non-radiative hydrodynamics. The comoving gravitational
softening length was fixed to $\epsilon=5.0\,h^{-1}\,\mathrm{kpc}$
(Plummer-equivalent).

For the present work, we use the output of the re-simulation at
redshift $z=0.297$. We separately analyse six sub-boxes, each of them
centred on one of the six most massive halos. The comoving side-length
of each sub-box was set to $\sim14.45\,h^{-1}\,\mathrm{Mpc}$,
corresponding to $1^\circ$ on the observer's sky.

\subsection{Weak-lensing simulations}

The particles in each sub-box sample a three-dimensional density
field. Their positions are interpolated on a grid of $512^3$ cells
using the {\em Triangular Shaped Cloud} method \citep{HO88.1}. Then,
the three-dimensional density field is projected along its coordinate
axes, thus producing three surface density maps, which we use as
independent lens planes in the following lensing simulations.

The lensing simulations are performed by tracing a bundle of
$2048\times2048$ light rays through a regular grid, covering the
central quarter of the lens plane, corresponding to a field of view of
$1^\circ$. The deflection angles $\vec\alpha_{\rm sig}(i,j)$ of each
ray $(i,j)$ are computed according to the method described in several
earlier papers \citep[see e.g.][]{ME00.1,ME01.1,ME04.1}.

Weak gravitational lensing by large-scale structures between the
observer and the sources is included as follows. From the power
spectrum (\ref{eq:lss}) of the effective convergence, we obtain the
power spectrum of the effective lensing potential,
\begin{equation}
  P_\psi(k)=\frac{4}{k^2}P_\kappa(k)\;.
\label{eq:ppsi}
\end{equation} 
Since we only aim at measuring the efficiency of our filter for
detecting a shear signal, the source redshift distribution is not
important. Thus, all sources are assumed to be at the same redshift
$z_\mathrm{s}$, considering the two cases $z_\mathrm{s}=1$ or
$z_\mathrm{s}=2$. The upper integration limit in Eq.~(\ref{eq:klss})
is then the comoving distance to the source plane, $w_{\rm s}=w(z_{\rm
s})$, and the distance distribution of the sources is a delta
function,
\begin{equation}
  G(w)=\delta(w-w_{\rm s})\;.
\end{equation}

We generate the effective lensing potential in Fourier space as a
Gaussian random field. Using Eq.~(\ref{eq:alpha}), we find the
large-scale structure contributions to the deflection angles,
$\vec{\alpha}_{\rm LSS}(i,j)$. The total deflection angle for each ray
$(i,j)$ is then
\begin{equation}
  \vec\alpha_{\rm tot}=\vec\alpha_{\rm sig}+\vec\alpha_{\rm LSS}\;.
\end{equation} 
Reduced-shear maps for each field are straightforwardly derived from
the deflection angle as explained in Sect.~\ref{sect:wlgeneral}.

Background galaxies are randomly placed and oriented, and their
ellipticities drawn from the distribution
\begin{equation}
  p_{\rm s}(|\epsilon_\mathrm{s}|)=
  \frac{\exp\left[
    \left(1-|\epsilon_\mathrm{s}|^2\right)/\sigma_{\epsilon_\mathrm{s}}^2
  \right]}{\pi\sigma_{\epsilon_\mathrm{s}}^2\left[
    \exp\left(1/\sigma_{\epsilon_\mathrm{s}}^2\right)-1
  \right]}
\end{equation}
with $\sigma_{\epsilon_\mathrm{s}}=0.3$. We assume a background galaxy
number density of $n_g=25\,\mathrm{arcmin}^{-2}$. Therefore, each field
contains $\sim90,000$ galaxies. ``Observed'' ellipticities are
obtained from Eq.~(\ref{eq:ell}), and a catalogue containing the
shapes and the positions of all sources in each field is compiled for
the following analysis.

\section{Results}

\subsection{Applying the filter}

We now describe how we analyse the simulated catalogues. The integral
in Eq.~(\ref{eq:estimate}) is approximated by a sum over galaxy
images. Since the ellipticity $\epsilon$ is an estimator for
$2\gamma$, we can write
\begin{equation}
  A_{\rm est}(\vec\theta)=\frac{1}{n_g}\,
  \sum_i\epsilon_{{\rm t}i}(\vec\theta_i)\,
  \Psi(|\vec\theta_i-\vec\theta|)\;,
\label{eq:discrsig} 
\end{equation}
where $\epsilon_{\mathrm{t}i}(\vec\theta_i)$ denotes the tangential
component of the $i$th galaxy's ellipticity relative to the line
connecting $\vec\theta_i$ and $\vec\theta$. Thus, the filtering is
performed in the real domain, which requires to Fourier back-transform
the filter function from Eq.~(\ref{eq:optfilter}).

We measure $A_{\rm est}$ on a regular grid of $128\times128$ points
covering the field. The noise estimate in $A_{\rm est}$ is obtained
from
\begin{equation}
  \sigma^2\left[A_{\rm est}(\vec\theta)\right]=
  \frac{1}{2N^2}\,\sum_i|\epsilon_{{\rm t}i}(\vec\theta_i)|^2 
  \Psi^2(|\vec\theta_i-\vec\theta|)\;.
\label{eq:discrnoise}
\end{equation}
We produce maps of the signal-to-noise ratio ($S/N$) for all our
fields. We then search for peaks with heights above a minimal $S/N$.

\begin{figure}[ht]
  \includegraphics[width=\hsize]{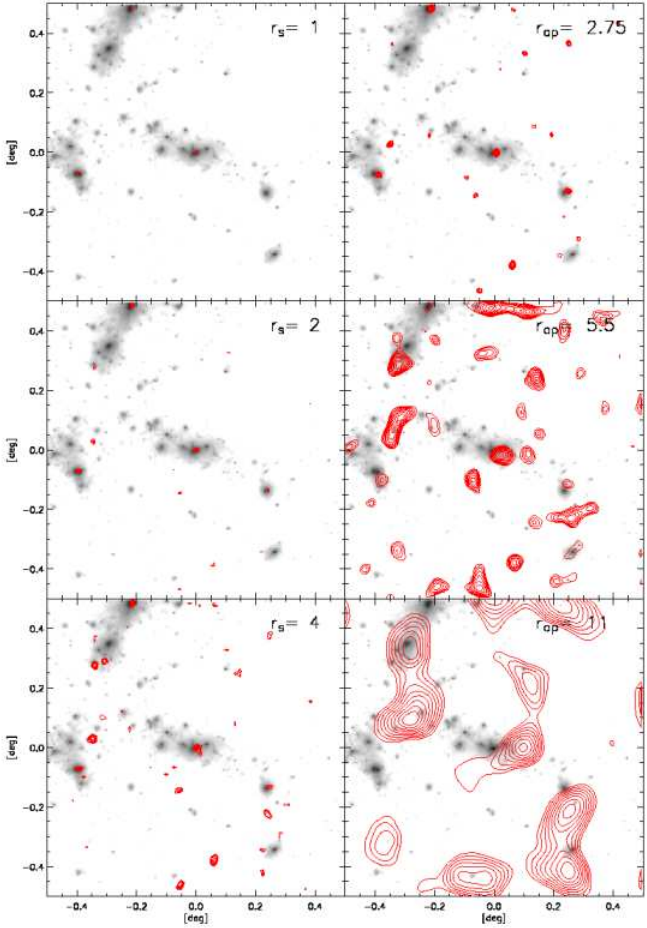}
\caption{Signal-to-noise contours above $5\sigma$ for different filter
  scales, assuming sources at redshift $z_\mathrm{s}=2$. Results
  obtained by applying the optimal filter and the aperture mass are
  shown in the left and right panels, respectively. From top to
  bottom, the scales are $1'$, $2'$ and $4'$ for the optimal filter,
  and $2.75'$, $5.5'$ and $11'$ for the aperture mass. The contours
  are overlaid in all panels on the surface-density map in the field
  centered on the third most massive halo in our simulation.}
\label{fig:3}
\end{figure} 

For comparison, similar signal-to-noise maps are produced using the
aperture mass statistic. In this case, we also use
Eqs.~(\ref{eq:discrsig}) and (\ref{eq:discrnoise}) for estimating the
signal and the noise, but we substitute the filter function $\Psi$
with the function $Q$ discussed in Sect.~\ref{sect:map}.

\subsection{Signal-to-noise maps}

We show in this section how the filter performs on one projection of
the field centred on the third most massive halo in the simulation
box. The results for the fields centred on the other five halos are
qualitatively and quantitatively very similar. The simulation we use
represents a fairly complex case to study. Indeed, the field is
crowded with $\sim15$ subhalos of different masses and some
filamentary structure connecting them.

Both the optimal filter and the aperture mass are designed for the
optimal detection of halos of a given scale. For the optimal filter,
this can be set to the scale radius of the NFW profile. However, this
definition does not apply to the aperture mass. In fact, the size of
the aperture is defined as the radius where the filter function $Q$
has its first root. In order to compare the efficiency of the two
filters for the detection of the same class of object, we have
empirically adjusted the size of the aperture such that it resembles
the scale of the signal profile used in the optimal filter. 

We show in Fig.~\ref{fig:3} the signal-to-noise contours for
detections reaching above $5\sigma$. The left and the right panels
refer to the aperture mass and the optimal filter, respectively. From
top to bottom, the filter scale is $1'$, $2'$, $4'$ for the optimal
filter, corresponding to scales of $2.75'$, $5.5'$ and $11'$ for the
aperture mass. In order to allow the reader to see which detections
are real and which are spurious, we overlay the contours on
surface-density maps from the numerical simulation. The halos which
are identified by the halo finder are marked by circles with radii
proportional to their virial radii.

\begin{figure}[ht]
  \includegraphics[width=\hsize]{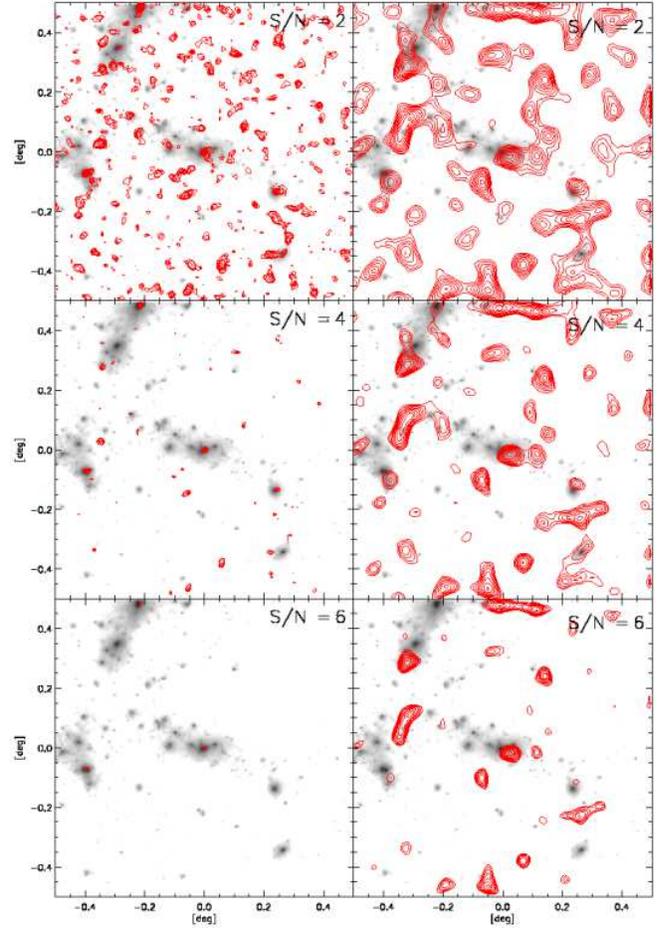}
\caption{Signal-to-noise contours for a fixed filter scale of $2'$ for
  the optimal filter (left panels) and of $5.5'$ for the aperture
  mass (right panels) for sources at redshift $z_\mathrm{s}=2$. From
  the top to the bottom panel, contours start at $2\sigma$, $4\sigma$
  and $6\sigma$, respectively. In all panels, the contours are
  overlaid on the surface-density map in the field centered on the
  third most massive halo in our simulation.}
\label{fig:4}
\end{figure} 

The optimal filter performs much better than the aperture mass in
detecting the dark matter concentrations present in our simulated
field. On all scales, the signal-to-noise ratios are larger for the
optimal filter than for the aperture mass.

Spurious detections occur with both filters. A further inspection
shows that these false matter concentrations are in fact connected to
over-densities produced by the projection of the large-scale
structures along the line-of-sight. As discussed in
Sect.~\ref{sect:noises}, these misidentifications cannot strictly be
avoided, for the large-scale structure power spectrum is still large
on the spatial scales where most of the signal resides. A possible
method for discriminating between true halos and large-scale structure
superpositions could be to group the sources in redshift bins. In
fact, the large-scale structure pattern is expected to change for
sources at different redshifts. This method is currently being tested
using numerical simulations (Meneghetti et al., 2005, in preparation).

It is remarkable that large-scale structure contaminations are much
less important for the optimal filter compared to the aperture mass,
especially when aiming at detecting clusters of large mass and low
redshift, having large scale radii.

In Fig.~\ref{fig:4}, we show the detections above a minimal
signal-to-noise ratio for a fixed filter scale of $2$ arc minutes for
our filter (right panels) and of $5.5$ arc minutes for the aperture
mass (left column). From the top to the bottom panels, the contours
start at $2$, $4$ and $6\sigma$, respectively. Again, the efficiency
of the optimal filter is remarkably higher than that of the aperture
mass. No detections above $3\sigma$ are found with $M_{\rm ap}$, while
several true peaks are found using the optimal filter at more than
$6\sigma$.

\begin{figure}[ht]
  \includegraphics[width=\hsize]{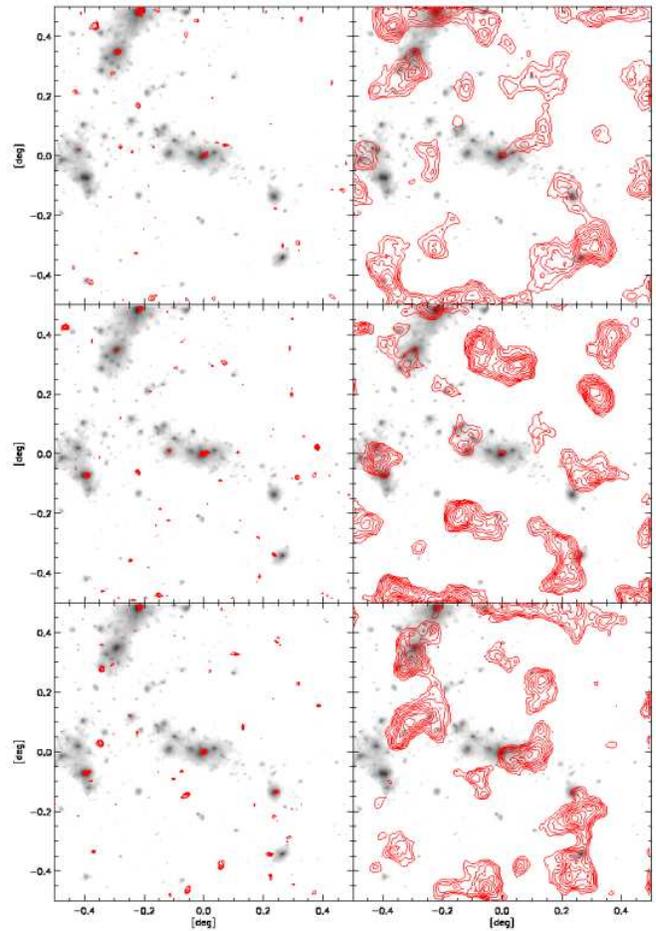}
\caption{Signal-to-noise contours above $4\sigma$ for different
  realizations of the large-scale structure. Results obtained by
  applying the optimal filter with and without filtering out the
  large-scale structure are shown in the left and in the right panels,
  respectively. The scale of the filter is $2'$ and the source
  redshift is $z_\mathrm{s}=2$.}
\label{fig:5}
\end{figure}

\begin{figure}[ht]
  \includegraphics[width=\hsize]{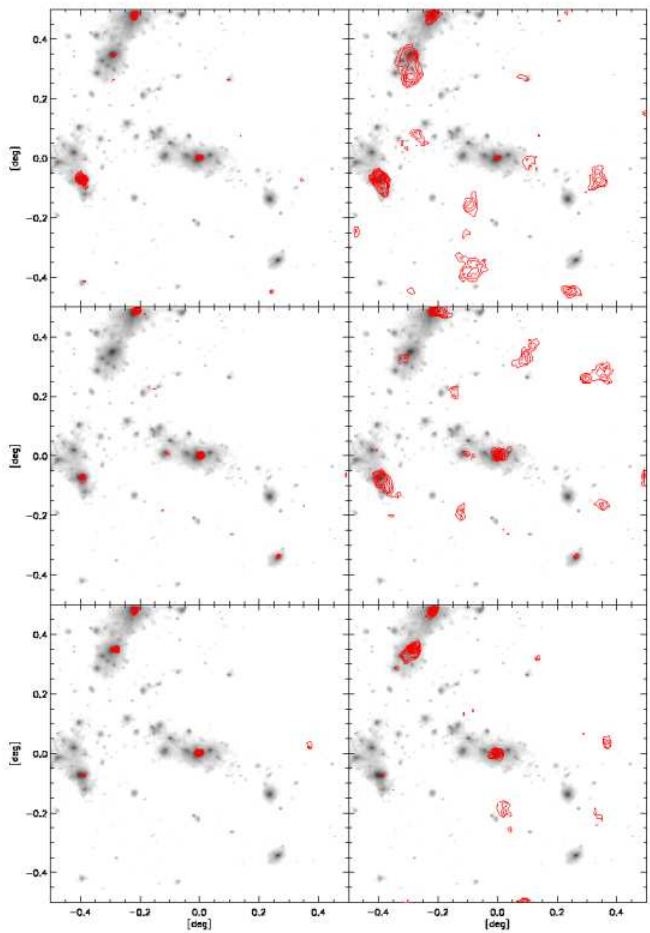}
\caption{As Fig.~\ref{fig:5}, but with the sources placed at redshift
  $z_\mathrm{s}=1$.}
\label{fig:6}
\end{figure}

Apparently the chance of a halo to be detected does not depend only on
its mass or on the amplitude of the shear which it is able to
produce. Indeed, the noise produced by the large-scale structure can
locally disturb the signal, such that even a relatively large halo can
fall below detectability due to the local effective shear produced by
the large-scale structure. As a simple test, we perform several
weak-lensing simulations keeping the background galaxy population
fixed and changing the large-scale structure in front of them, by
using different realisations of the Gaussian random deflection-angle
field. The results are shown in the sequence in Fig.~\ref{fig:5} for
sources redshift $z_\mathrm{s}=2$, and in Fig.~\ref{fig:6} for
$z_\mathrm{s}=1$. We use a filter scale of $2$ arc minutes and plot
the signal-to-noise contours above $4\sigma$ in all panels. In the
right panels of images, the optimal filter is applied in order to
remove the large-scale structure contamination. Clearly, some of the
clumps which are detected for a given realization of the large-scale
structure, are entirely missed when a different seed is used. For
better illustrating the influence of the large-scale structures on the
detection efficiency, we show in the left panels the same detections
without including the large-scale structure noise in the construction
of the optimal filter. Spurious halo detections are strongly reduced
by using low-redshift sources because of the weaker large-scale
structure contribution along the line-of-sight. This causes the
signal-to-noise ratio to increase despite the clusters' being less
efficient lenses for the lower-redshift sources. Compared to
Fig.~\ref{fig:5} the clusters' lensing efficency is smaller because of
the lower sources distance, but, for the same reason, there is a
strong reduction in the large-scale structure contamination which
leads to a gain in the signal-to-noise ratio.

\section{Summary and discussion}

We have developed a linear filter for detecting the weak-lensing
signal of dark-matter halos. Given the expected shape $\tau$ of the
halo signal, the filter is uniquely determined by the two requirements
that it be unbiased and maximise the signal-to-noise ratio. The filter
shape is modified by the noise power spectrum, into which we include
the shot noise from the random positions of the galaxies used to
measure the gravitational shear, the noise from their random intrinsic
ellipticities, and the weak-lensing signal contributed by the
large-scale structure. Since the total noise power spectrum is
monotonically falling over the scale of the shape function $\tau$, the
inclusion of the noise narrows the filter considerably.

Our primary goals are to obtain a filter which separates halos from
spurious peaks in the lensing effects of the projected large-scale
structure, and whose results are stable against small changes in the
filter scale. Our motivation is that applications of another statistic
used for detecting dark-matter halos, the aperture mass
$M_\mathrm{ap}$, tend to detect large numbers of dark peaks,
i.e.~peaks in the lensing signal which are not found to correspond to
detectable signal in the optical or X-ray wave bands. Moreover, these
peak detections tend to be unstable against changes in the
characteristic scale of $M_\mathrm{ap}$, i.e.~the significance of
their detection may change wildly upon even small changes of scale.

It is clear that a clean separation of halos from large-scale
structure is impossible because large-scale structure can be
considered as being composed of dark-matter halos. We thus choose the
operational definition that the halos we wish to detect are
distinguished from large-scale structure as being objects on scales
smaller than the non-linear scale. Consequently, we model the
large-scale structure ``noise'' in the weak-lensing signal as being
due to linearly evolved structures alone.

We perform numerical simulations for testing our new filter. The
results, summarised in Figs.~\ref{fig:3}, \ref{fig:4} and \ref{fig:5},
demonstrate that the filter meets all of our goals. Changes in the
filter scale leave the signal much more stable than the aperture mass
does, as Fig.~\ref{fig:3} shows. Fig.~\ref{fig:4} illustrates that the
signal-to-noise ratio produced with our filter is typically
substantially higher than that of the aperture mass, and
Fig.~\ref{fig:5} demonstrates that filtering halos against large-scale
structure lensing is indeed very efficient. Compared to the aperture
mass, the number of spurious peaks is also substantially reduced.

We thus propose this filter as an alternative statistic for detecting
dark-matter halos. Of course, the aperture mass remains a very
valuable statistic which has its significant strengths in analyses of
lensing by large-scale structures. It appears, however, from our
numerical simulations that a fair fraction of the peaks found with the
aperture mass may in reality not be caused by halos, but by peaks in
the projected large-scale structure.

\acknowledgements{We are grateful to Catherine Heymans and Michael
  Schirmer for helpful discussions. The $N$-body simulations were
  performed at the ``Centro Interuniversitario del Nord-Est per il
  Calcolo Elettronico'' (CINECA, Bologna), with CPU time assigned
  under an INAF-CINECA grant.}


\end{document}